# β-Ga$_2$O$_3$ Double Gate Junctionless FET with an Efficient Volume Depletion Region

Dariush Madadi, *IEEE*, Ali A. Orouji, *Senior Member, IEEE*

*Abstract—* This paper presents a new β-Ga$_2$O$_3$ junctionless double gate Metal-Oxide-Field-Semiconductor-Effect-Transistor (βDG-JL-FET) that a P$^+$ packet embedded in the oxide layer (PO-βDG-JL-FET) for high-voltage applications. Our goal is to achieve an efficient volume depletion region by placing a P$^+$ layer of silicon. We show that the proposed structure has a subthreshold swing ~ 64 mV/decade and it suppressed the band to band tunneling (BTBT) phenomenon. Also, the PO-βDG-JL-FET structure has a high $I_{on}/I_{off}$ ~ 1.3×10$^{15}$. The embedded layer reduces the off-current ($I_{OFF}$) by ~ 10$^{-4}$, while the on-current ($I_{ON}$) reduces slightly. Besides, we show that the proposed structure has acceptable $I_{off}$ value in a range of gate work functions which help us to the optimization of designs in terms of area, power gain, and leakage current. The leakage current of the proposed structure is ~ 7 × 10$^{-17}$ A in 400 K temperature. Furthermore, the fabrication process steps of the proposed structure will be investigated.

## I. Introduction

Metal-Oxide-Semiconductor-Field-Effect-Transistors (MOSFETs) have been investigated, and there was high interest among the researchers to investigate their electrical performances and scaling of MOSFETs has been reached its limit due to several problems such as the increment of short-channel-effects, difficulty in define the boundary between the source and drain regions in doping profiles, and the large leakage currents ($I_{off}$) [1][2][3][4], [5][6] [7]. The structures like double gate, and Fin MOSFETs improve the short-channel-effects and the $I_{off}$ current behavior but difficulty in defining the boundary between the source and drain regions in doping profiles still exists [8]. The junctionless FET (JL-FET) that have no junction at the source–channel and channel–drain interfaces provides a solution to this problem but they show poor high-voltage characteristics[9]–[14][15], [16]–[21][22][23]. Characteristics of high-voltage breakdown have been investigated in multigate structures[24] [25]and junctionless FETs[26]. By using high-wide bandgap semiconductors such as beta gallium oxide (β-Ga$_2$O$_3$) we can overcome the poor high-voltage characteristics to improve the characteristics of JL-FET in high-voltage applications. Beta phase gallium oxide (β-Ga$_2$O$_3$) is a semiconductor with a high bandgap and is a suitable choice for high-power switching and high-voltage RF electronics. β-Ga$_2$O$_3$ has a bandgap ($E_g$) of approximately 4.9 eV with the breakdown filed ~ ($E_c$) of 8 MV/cm. The breakdown field for β-Ga$_2$O$_3$ is 2 - 3 times higher than materials such as silicon carbide (SiC) or gallium nitride (GaN) [27]–[33]. Various studies focused on the junctionless structures but the application of β-Ga$_2$O$_3$ for JL-FETs has not yet been reported.

In this paper, we propose a JL-FET structure by an embedded P$^+$ layer and with the β-Ga$_2$O$_3$ material to improve the off-current and the characteristics of high-voltage due to an efficient volume depletion region and suppressing the band to band tunneling (BTBT). By using a wide bandgap semiconductor, tunneling width at the interface of channel-drain will change and this phenomenon affects the BTBT and leakage current. Thus, the β-Ga$_2$O$_3$ JL-FET structure mitigates the complexity of using the FinFETs or gate-all-around structures to achieve a higher breakdown in JL-FETs, and by using the simple structure we can obtain better results. By using JL-FETs, and due to the current conduction mechanism, in the partial depletion mode, the current flows farther from the surface unlike the conventional MOSFETs and higher mobility can be obtained. Besides, the proposed structure has a superior subthreshold slope (SS) ~ 64 mV/decade and the off current, thus making the structure very suitable for power operations. Unlike the conventional MOSFETs, the counter doping and additional annealing steps may not be required in β-Ga$_2$O$_3$ JL-FETs to neutralize the interface traps[34].

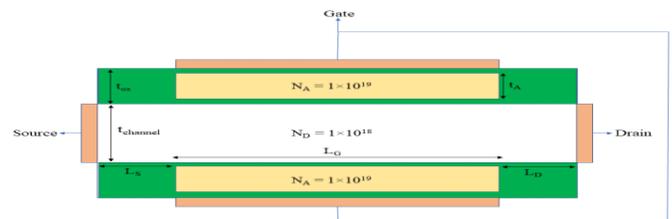

Fig. 1. Schematic view of β-Ga$_2$O$_3$ JL-FET.

## II. Proposed Structure and Simulation Models

Fig. 1 shows the view of β-Ga$_2$O$_3$ JL-FET in 2 Dimensional. All the simulations of structures are done with SILVACO TCAD software[35]. Important parameters of the proposed structure are in Table. 1 and all of the parameters of both the structures are the same except the embedded layer in the proposed structure at the oxide of the gate. The material of the active regions of the device is β-$Ga_2O_3$ and we use the silicon material for the embedded layer. Also, the length of the channel and the embedded layer is the same. Besides, channel, source, and drain regions have uniform doping density with the same doping concentration. All simulations are under a similar situation and at 26.85 Celsius (300 Kelvin) temperature. The crystalline orientation is in ⟨100⟩ direction and thermal conductivity at room temperature is set at 12 W (mK)$^{-1}$. Due to interface charges and defects traps density between $SiO_2$ and $Ga_2O_3$, ionized impurity scattering (Coulomb scattering) is considered that causes the lower mobility in the channel[4]. Also, surface roughness scattering and phonon scattering mechanisms are considered to investigate effects on mobility. All scattering mechanisms are considered in ATLAS for investigating the mobility of both the structures. Interface charge density is considered at $5.5 \times 10^{12}$ cm$^{-2}$[36]. Thermal contact is considered in both the devices for consideration of thermal effects. Proper models for the mobility and leakage current are set. The velocity saturation is set to $2 \times 10^7$ m/Sec and for consideration of the velocity saturation model for non-silicon semiconductors such as the β-$Ga_2O_3$, statement of $V_{sat} = 2 \times 10^7$ m/Sec is considered. Fermi Dirac, SRH recombination, BGN, and BTBT models were used. Also, The β-$Ga_2O_3$/oxide interface trap density (Dit) is set by considering the fixed charge traps ($2.3 \times 10^{12}$ cm$^{-2}$) and acceptor traps ($2.7 \times 10^{12}$ cm$^{-2}$) at the β-$Ga_2O_3$–$SiO_2$ interface[37][36]. Simulation results are calibrated with experimental results [38] to validate the accuracy of simulation models in the structures and the output characteristic of the proposed structure is shown in Fig. 2.

Table 1. Important parameters of the proposed structure in our work.

| The device parameters are used in the simulations | Value |
|---|---|
| Thickness of oxide ($t_{ox}$) | 6 nm |
| Channel length ($L_G$) | 1 μm |
| Source length ($L_S$) | 100 nm |
| Drain length ($L_D$) | 100 nm |
| Embedded P$^+$ Layer thickness ($T_A$) | 4 nm |
| β-$Ga_2O_3$ layer thickness ($t_{ch}$) | 10 nm |
| β-$Ga_2O_3$ doping ($N_D$) | $1 \times 10^{18}$ cm$^{-3}$ |
| Embedded P$^+$ Layer doping ($N_A$) | $1 \times 10^{19}$ cm$^{-3}$ |
| Material of gate oxide | $SiO_2$ |
| Gate work function | 4.7 eV |

### III. DC RESULTS AND DISCUSSIONS

In this paper, we propose a β-$Ga_2O_3$ junctionless double gate Metal-Oxide-Semiconductor-Field-Effect-Transistor with a P$^+$ packet that embedded in the oxide layer (PO-βDG-JL-FET) for achieving an efficient volume depletion region and by embedding this layer, the electron concentrations in the structure reduce at off-state mode ($V_{gs}$ = 0 V, $V_{ds}$ = 0.5 V). Our goal from the embedding P$^+$ layer is to achieve a full depletion region from the carriers in the off-state mode. In our work, the embedding P$^+$ layer prevents the flow of the majority of carriers from the source to the drain region. The current flows from the middle of the structure unlike the conventional MOSFET that the current flows near the β-$Ga_2O_3$/$SiO_2$ surface, and the proposed structure has better on-current than the conventional MOSFETs. By increasing the depletion region at the off-state, the flowing of carriers will be fewer, and the conductivity of the channel decreases.

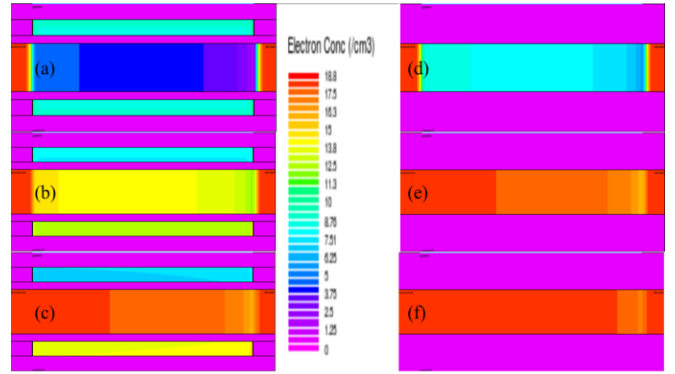

Fig. 3. The electron concentrations at $V_{ds}$ = 0.5 V of the βDG-JL-FET in (a) $V_{gs}$ = 0 V, (b) $V_{gs}$ = 0.5 V, (c) $V_{gs}$ = 1 V and PO-βDG-JL-FET in (d) $V_{gs}$ = 0 V, (e) $V_{gs}$ = 0.5 V, (f) $V_{gs}$ = 1 V at logarithm scale.

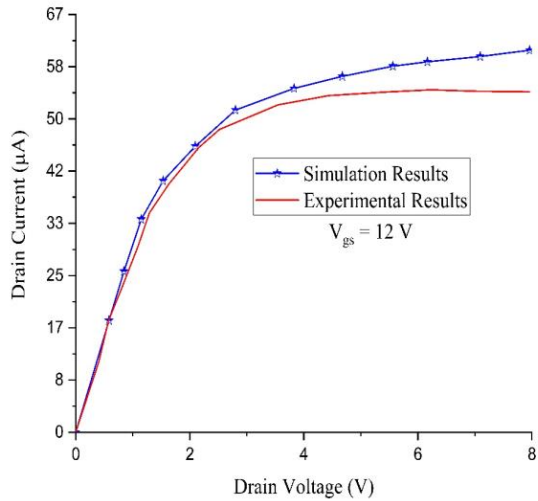

Fig. 2. Output characteristics calibrated by experimental results[38].

In the comparison between Fig. 3 (a) and (d) in the $V_{gs}$ = 0 $V_{ds}$ = 0.5 V, we can see that the proposed structure has a higher depletion region due to the embedded P$^+$ layer versus the βDG-JL-FET at off-state mode. Also, at the partial depletion-mode in Fig. 3 (b) and (e), and because of the idea that we added to the structure, improvement of the depletion region in the proposed structure, and lower leakage current than the βDG-JL-FET will be seen.

Fig. 3 shows the electron concentration contour at the off-state mode and due to the P$^+$ layer in the proposed structure, the electrons in the channel will disappear and because of an efficient volume depletion region added to our work, at the top and bottom of the structure, the channel will be completely depleted from the carriers in the proposed structure in the off-state mode while the βDG-JL-FET is not fully depleted and the electron concentration of channel has a higher value versus the proposed structure. For a better view of the effect of the embedded layer, we show the electron concentration at various $V_{gs}$ in Fig. 4 by the cut line in the lateral position.

As we can see in the figure, by increasing $V_{gs}$, the electron concentration increases, and as a result, the depletion region of structures decreases, and the leakage current ($I_{OFF}$) increases. At the $V_{gs} = 0$ V, the electron concentrations of the proposed structure have a better value ~ 4 times versus the βDG-JL-FET.

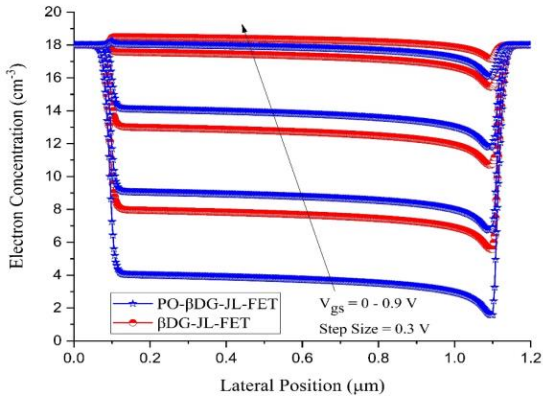

Fig. 4. The electron concentrations of both the structures at $V_{ds} = 0.5$ V and various $V_{gs}$ at the logarithm scale.

Fig. 5 shows the transfer characteristics (I-V) of both the structures in the linear and logarithm scale at $V_{gs} = 1$ V and $V_{ds} = 0.5$ V and the gate work function = 4.8 eV. The cut line is 3 nm under the gate oxide in the lateral position. As mentioned, due to the embedded layer and because of the creation of efficient volume depletion in the proposed structure, the leakage current improves. The Embedded P$^+$ layer improves the off-current ~ 4 times, while the on-current decreases a bit.

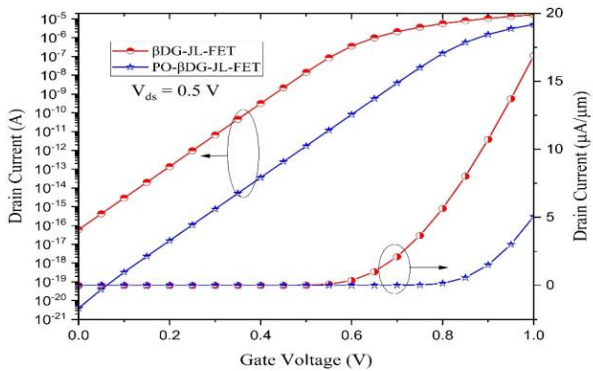

Fig. 5. Transfer characteristics of both the structures in the linear and logarithm scale.

The gate voltage at which the neutral place at the middle of the channel vanishes because of the combining of depletion area widths contributed through the gate contact is described for the voltage threshold of the JL FETs. We can see the threshold voltage formula in JLFETs[15]:

$$V_{TH} = V_{FB} - \frac{qN_D t^2_{Ga_2O_3}}{2\varepsilon_{Ga_2O_3}} - \frac{qN_D t_{Ga_2O_3} t_{ox}}{\varepsilon_{ox}} \quad (1)$$

Where $t_{ox}$ and $t_{Ga2O3}$ are the gate oxide and $Ga_2O_3$ thickness and $\varepsilon_{ox}$ and $\varepsilon_{Ga2O3}$ are the permittivity's of $SiO_2$ and $Ga_2O_3$ material. Also, it should be noted that $q$ is the electronic charge and $N_D$ is the donor carriers and $V_{FB}$ is the voltage in the flat-band condition. The expansion of depletion in the channel region increases because of the embedded layer with a growth of the effective channel area. Consequently, control through the gate dictates the threshold voltage of the proposed structure that increases the effective channel area results in a lower leakage current due to a growth in the threshold voltage that can be observed in Fig. 5. Besides, due to the higher depletion region of the proposed structure, the PO-βDG-JL-FET structure has a higher barrier in the channel-source side and as a result, effect of DIBL (drain induced barrier lowering) improves in comparison the βDG-JL-FET.

The other goal is to use β-$Ga_2O_3$ that is a semiconductor with an ultra-wide bandgap ($E_g$) that can solve poor high-voltage characteristics in the JL-FETs.

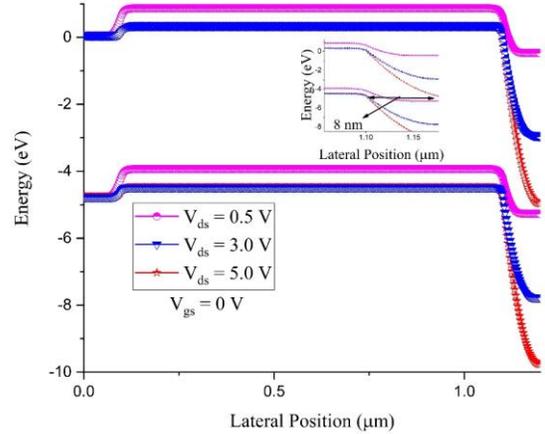

Fig. 6. Energy band of diagram for the proposed structure at $V_{gs} = 0$ V by the cut line in the lateral position.

We see from Fig. 6 that there is no overlap between the conduction and valance bands, and hence, BTBT does not occur at low drain voltages due to using β-$Ga_2O_3$ with a high bandgap, unlike the silicon JL-FETs. The cut line is 3 nm under the gate oxide in the lateral position. By increasing $V_{ds}$, and at $V_{ds} \geq E_g$, the conduction and valance bands start to overlapping and this phenomenon triggers the BTBT that causes to degrades off-current. Also, the tunneling width at the off-state condition in the proposed structure is 8 nm while this value in the βDG-JL-FET is 6 nm. As a result, the PO-βDG-JL-FET structure has a fewer leakage current than the βDG-JL-FET because of suppression BTBT and shows the improvement of high-voltage characteristics than the silicon JL-FET or the βDG-JL-FET.

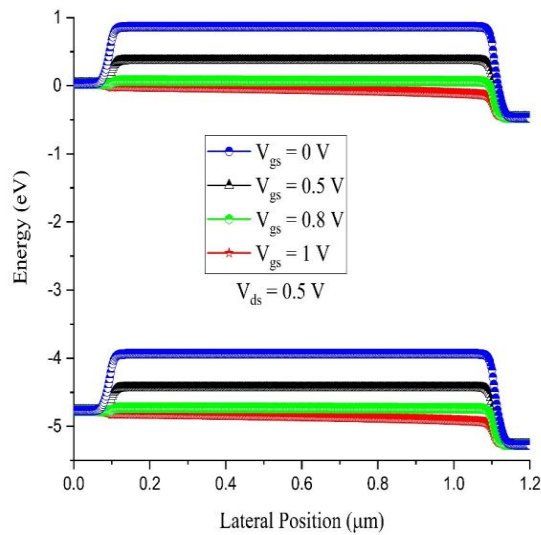

Fig. 6. Energy band diagrams for the proposed structure at different $V_{gs}$ and $V_{ds}$ = 0.5 V by the cut line in the lateral position.

Fig. 6 shows the energy band diagrams of for the proposed structure at different $V_{gs}$ and $V_{ds}$ = 0.5 V by cutline 3 nm below the top gate oxide. As we observed in the figure, the channel of the proposed is completely depleted from the carriers in the lower $V_{gs}$, and by increasing gate voltage, the energy bands will be flat at $V_{gs}$ = 1 V. In the off-state condition, while the channel is depleted due to the gate work function of the gate, the added $P^+$ layer creates an efficient volume depletion region in the interface and extends to the source and drain region. By increasing $V_{gs}$, the $P^+$ layer getting depleted from the P carriers, and due to the diffusion of electrons from the channel, the source and drain regions cause to invert the $P^+$ layer to the N-type layer. As a result, the added layer plays a good role in the special range (lower $V_{gs}$) and after inverting the $P^+$ layer to the N-type layer, the layer will cause problems for the proposed structure.

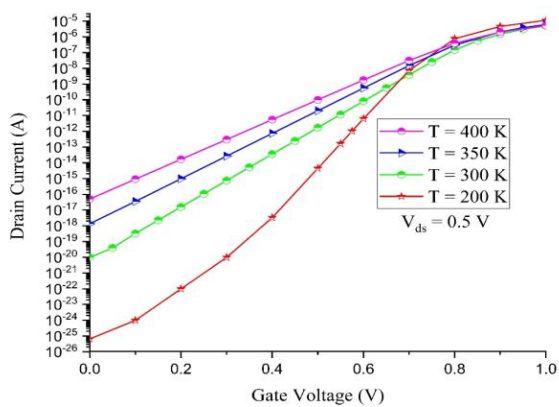

Fig. 7. Transfer characteristics of the proposed structure at various temperatures.

In Fig. 7, we see the transfer characteristics of the proposed structure at various temperatures. By increasing temperature, the leakage current increases due to the higher rate of recombination and intrinsic concentration. These effects cause to have fewer depletion regions and extra carriers will appear for current conduction. Also, at lower temperatures, the effect of traps and defects on characteristics of the proposed structure reduces, and the leakage current decreases.

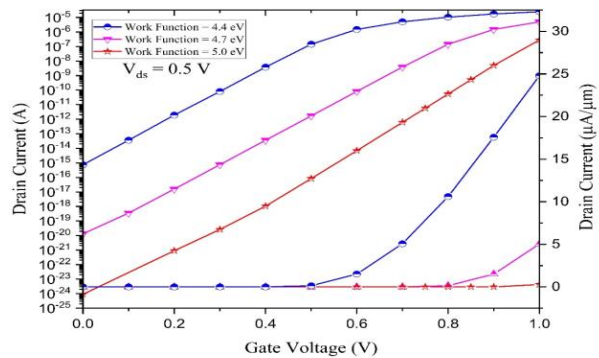

Fig. 8. Transfer characteristics of the proposed structure at various gate work functions.

Fig. 8 shows the transfer characteristics of the proposed structure at various gate work functions. As we see in the figure, by increasing the gate work function, the leakage current reduces but the on-current slightly reduces. We observed that at a various gate work function, still have a high $I_{on}/I_{off}$ ~ 1.3 × $10^{15}$ and due to this reason, we can have multiple threshold voltage and enables us to achieve the best designs. It notes that despite the restriction of silicon JL-FET at the short channel to have the gate work function of higher than 5.1 eV, the proposed structure can be operated at various of the gate work functions.

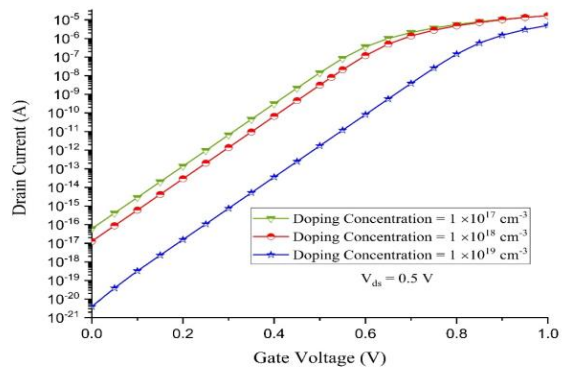

Fig. 9. Transfer characteristics of the proposed structure at various doping concentrations of the embedded layer.

Fig. 9 shows the I-V curve of the proposed structure at various carrier doping of the embedded layer. As can be seen, by increasing carrier doping, at the off-state mode, we have a higher depletion region due to the $P^+$ layer and the leakage current reduces versus the other concentrations. At the doping concentration = 1 × $10^{19}$ cm$^{-3}$ the leakage current improves ~ 4.5 times versus the doping concentration at = 1 × $10^{17}$ cm$^{-3}$. It should be noted that in junctionless FETs if the silicon layer is too extremely doped (higher than $1 \times 10^{19}$ cm$^{-3}$) especially in the center of the channel where the peak of doping concentration is located and/or too thin, it may be unfeasible to fully deplete the silicon layer from carriers and thus to switch off the device.

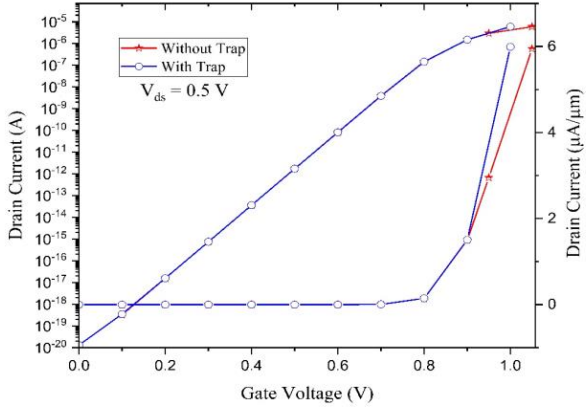

Fig. 10. Transfer characteristics of the proposed structure with and without interface traps in linear and logarithm scale.

In Fig. 10, we observe the transfer characteristics of the proposed structure at the logarithm scale with and without semiconductor/oxide interface traps in linear and logarithm scale. As can be seen in the figure, the β-Ga$_2$O$_3$ JL-FETs are not sensitive to the traps at the Ga$_2$O$_3$-SiO$_2$ interface due to the conduction mechanism, unlike the conventional MOSFETs that have conduction at the surface and as a result, the mobility of our structure improves. JL-FETs current conduction mechanism at $V_{gs} \leq V_{FB}$ (Flat band voltage) is in the middle of the channel and traps of the semiconductor-oxide interface will not affect on characteristics of the structure. However, in the accumulation mode (upper than $V_{FB}$), the depletion region in the channel disappears and the electron concentration at the surface will be more than the doped electron concentration.

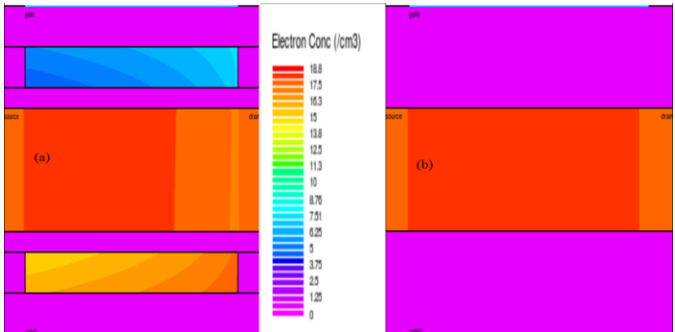

Fig. 11. The electron concentrations of (a) PO-βDG-JL-FET and (b) βDG-JL-FET in accumulation mode $V_{gs} = 1$ V and $V_{ds} = 0.5$ V at logarithm scale.

In accumulation mode, unlike the conventional β-Ga$_2$O$_3$ MOSFET, after situating of free electrons by semiconductor-oxide traps, the JL-FET proposed structure has enough free electrons to have high on-current and plays a role as conventional MOSFETs. As can be seen from the figure, the PO-βDG-JL-FET structure has a lower electron concentration and as a result, the electric field of the proposed structure will be lower than the βDG-JL-FET.

## IV. AC RESULTS AND DISCUSSIONS

We investigated the various parameters in the dc mode that we obtained superior results in the proposed structure, now we want to investigate the ac mode characteristics for both structures. Fig. 12 shows the capacitance between the gate and drain side ($C_{gd}$) and capacitance between the gate and source side ($C_{gs}$) at different frequencies in both structures.

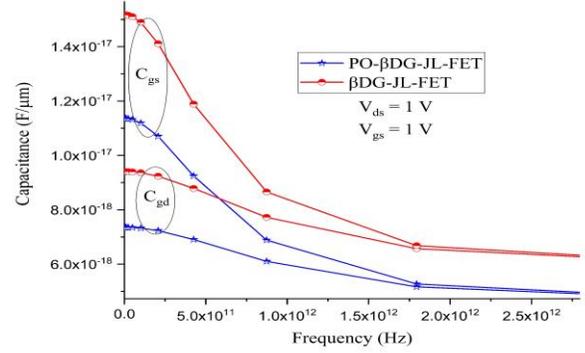

Fig. 12. $C_{gd}$ and $C_{gs}$ at different frequencies in both the structures.

Important reason for this phenomenon, relying on the difference in the gate oxide thickness due to the embedding silicon layer in the oxide layer. Additionally, the other cause for reducing parasitic capacitance is due to an embedded P$^+$ layer that the depletion region of the channel extending into the source and drain areas and it causes to mitigate the electrical coupling among the gate-drain, and as a result, reduction of the gate-drain capacitance and gate-source capacitance will occur.

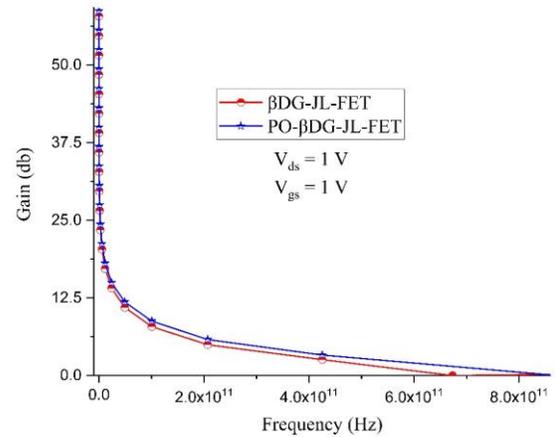

Figure. 13. Transducer power gain as a function of the frequency for both the structures.

The embedded P$^+$ layer with silicon that has a higher thermal conductivity versus β-Ga$_2$O$_3$ causes to have a lower lattice temperature. Fewer lattice temperature causes to have a higher power gain. Due to the enhancement of the power gain, the proposed structure is suitable for high-voltage and high-frequency applications. Fig. 13 shows transducer power gain as a function of the frequency for both the structures. As can be seen in the figure, because of the lower lattice temperature of the proposed structure, the gain of the PO-βDG-JL-FET structure is higher than βDG-JL-FET.

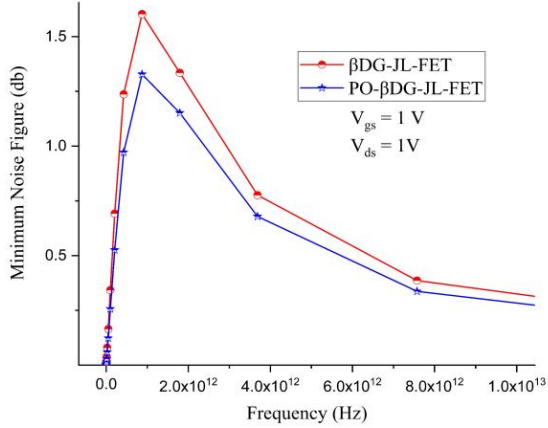

Fig. 14. Minimum noise figure as a function of frequency.

The minimum noise figure ($F_{min}$) is one of the important parameters for high-voltage and high-frequency applications. The $F_{min}$ of the structure is taken as follows[39][40]:

$$F_{min} = 1 + 5\pi f C_{gs} \sqrt{\frac{R_S + R_G}{g_m}} \qquad (2)$$

Which $R_S$ is the source resistance, $R_G$ is the gate resistance, $C_{gs}$ is gate-source capacitance and $g_m$ is the transconductance. As we discussed before, by fewer $C_{gs}$, $R_S$, and $R_G$, the minimum noise figure of the proposed structure has a better value than βDG-JL-FET.

V. FABRICATION PROCESS FOR THE PROPOSED STRUCTURE

Fig. 15 shows the steps of a process flow for the fabrication of the PO-βDG-JL-FET structure. The process of fabrication is divided into two wafers that have alike processing steps and finally are bonded together. We considered 5 nm for each wafer with an N-type β-Ga$_2$O$_3$ on the BOX layer with the SiO$_2$ layer, and the doping of the wafer is uniform with $1\times10^{19}$ cm$^{-3}$ concentration that is covered with a SiO$_2$ gate oxide as can be seen in Fig. 15 (a). Next, the intended part of the SiO$_2$ gate oxide replaced by a silicon material as mentioned at the following fabrication levels. In Fig. 15 (b), SiO$_2$ gate oxide is selectively etched by using buffered hydrogen fluoride (BHF) solution. Etching processes will occur at specific lengths and the other region of the wafer is protected by a photoresist mask. After etching processes, the photoresist was removed by warm acetone. Fig. 15 (c) shows the deposition of the silicon layer. Second, as exhibited in Fig. 15 (d), the atomic layer deposition of a SiO$_2$ is performed by using normal photolithography. Besides, the gate electrode contact is deposited. For reaching this goal, deposition by e-beam evaporation without annealing is done by utilizing the lift-off process [41]. In Fig 15 (e), the BOX layer is completely etched. It is noted that all process is similar for the other wafer. Then, as we see in Fig. 15 (f) and (g), two wafers are bonded together to have a complete structure. At the end of the process, the source and drain contacts can be fabricated by using e-beam deposition and annealing as can be seen in Fig. 15 (h).

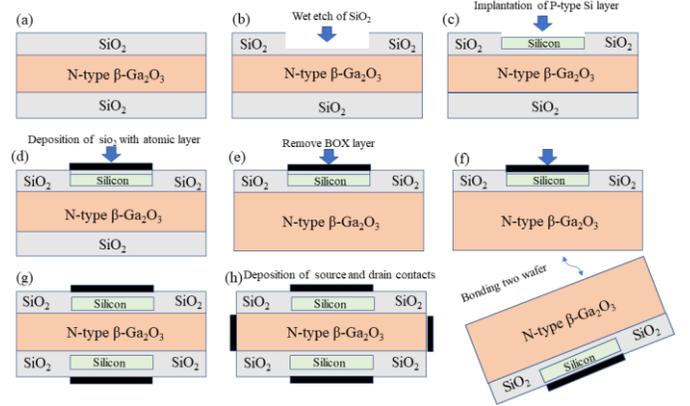

Fig. 15. The fabrication process for the PO-βDG-JL-FET structure. (a) growth of SiO$_2$, (b) wet etching of the top SiO$_2$ gate oxide, (c) deposition of the silicon layer, (d) atomic layer deposition of SiO$_2$ and deposition of the gate contact, (e) remove the BOX layer, (f), bonding of two wafers (g), complete of the structure without drain and source contacts (h) Schematic view of the proposed structure by considering deposition of the source and drain contacts.

VI. CONCLUSION

In this paper, we show a β-Ga$_2$O$_3$ junctionless double gate Metal Oxide Field Semiconductor Effect Transistor (βDG-JL-FET) that a P$^+$ packet embedded in the oxide layer (PO-βDG-JL-FET) for achieving an efficient volume depletion region to use in the high-voltage applications. The added layer due to using ultra-wide bandgap semiconductor helps us to improve the high-voltage characteristics by suppressing the BTBT effect. Besides, because of the mechanism of conduction in JL-FETs that the carriers flowing through the middle of the channel, traps have no impact on the characteristics of the structure, and as a result, the leakage current reduces. Because of acceptable high $I_{on}/I_{off}$ and high leakage current, multiple threshold voltage will be achieved and it enables us to achieve the best designs. Also, the AC results of the proposed structure have superior results versus the βDG-JL-FET.

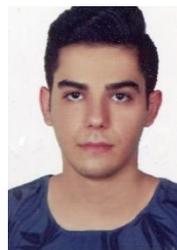

**Dariush Madadi** is currently pursuing Master of Science in micro and nanoelectronics in Semnan University. Madadi's research concentrates on semiconductors and nano electronics.

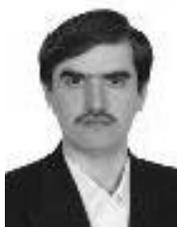

**Ali Asghar Orouji** (M'05–SM'14) received the Ph.D. degree from the IIT Delhi, Delhi, India, in 2006.
He has been a Faculty Member with Semnan University, Semnan, Iran, since 1992.